\documentclass[fleqn,usenatbib]{mnras}

\usepackage{mathptmx}
\usepackage{txfonts}

\usepackage[T1]{fontenc}

\DeclareRobustCommand{\VAN}[3]{#2}
\let\VANthebibliography\thebibliography
\def\thebibliography{\DeclareRobustCommand{\VAN}[3]{##3}\VANthebibliography}

\usepackage{graphicx}	
\usepackage{amssymb}	

\def \xmm {$XMM$-$Newton$}

\def \chandra {$Chandra$}
\def \suz {{\em Suzaku}}

\def \inte {$INTEGRAL$}

\def\epic{{EPIC}}
\def\pn{{pn}}
\def\mos{{MOS}}
\def\mosuno{{MOS1}}
\def\mosdue{{MOS2}}

\def \src {AX~J1714.1$-$3912}
\def \rxj {RX~J1713.7$-$3946}
\def \cxo {CXOU~J171343.9$-$391205}
\def \sgb {sgB[e]}

\def \nh {N$_\mathrm{H}$} 
\def \ergsec{\hbox{erg s$^{-1}$}}

\def \ferg {erg cm$^{-2}$ s$^{-1}$}
\def \hcm {\hbox {\ifmmode $ atom cm$^{-2}\else atom cm$^{-2}$\fi}}

\def \arcsec {\hbox{$^{\prime\prime}$}}

\def \apj {ApJ}
\def \aj {AJ}
\def \apjl {ApJL}
\def \apjs {ApJS}
\def \aap {A\&A}

\def \pasj {PASJ}

\def \mnras {MNRAS}

\def \ssr {Space Science Reviews}

\def\flux {\mbox{erg cm$^{-2}$ s$^{-1}$}}

\title[\xmm\ discovery of very high obscuration in \src]{\xmm\ discovery of very high obscuration in the  candidate Supergiant Fast X-ray Transient \src
\thanks{
Based on observations (ObsID 0804300901) obtained with {\em XMM-Newton}, an 
ESA science mission with instruments and contributions directly funded by ESA Member States and NASA.
}
}

\author[L. Sidoli et al.]{
{L.~Sidoli$^{1}$,\thanks{E-mail: lara.sidoli@inaf.it}}
V.~Sguera$^{2}$,
P.~Esposito$^{3,1}$,
L.~Oskinova$^{4,5}$
and M.~Polletta$^{1}$ 
\\
$^{1}$INAF, Istituto di Astrofisica Spaziale e Fisica Cosmica, Via A.\ Corti 12, 20133 Milano,  Italy  \\
$^{2}$INAF, Osservatorio di Astrofisica e Scienza dello Spazio, Via P.\ Gobetti 101, 40129 Bologna, Italy \\
$^{3}$Scuola Universitaria Superiore IUSS Pavia, Piazza della Vittoria 15, 27100, Pavia, Italy \\
$^{4}$Institute for Physics and Astronomy, University Potsdam, 14476 Potsdam, Germany  \\
$^{5}$Kazan Federal University, Kremlyovskaya 18, 420008 Kazan, Russia 
}
\date{Accepted 2022 March 9. Received 2022 January 25; in original form 2021 October 26}

\pubyear{2022}

\begin{document}
\label{firstpage}
\pagerange{\pageref{firstpage}--\pageref{lastpage}}
\maketitle

\begin{abstract}
  We have analysed an archival \xmm\ EPIC observation that serendipitously covered the sky position of a variable X-ray
  source \src, previously suggested to be a Supergiant Fast X-ray Transient (SFXT). During the \xmm\ observation
  the source is variable on a timescale of hundred seconds and shows two luminosity states, with a flaring activity
  followed by  unflared emission, with a variability amplitude of a factor of about 50.
  We have discovered an intense iron emission line with a centroid energy of 6.4 keV in the power law-like spectrum,
  modified by a large absorption (\nh$\sim$10$^{24}$~cm$^{-2}$), never observed before from this source. 
  This X-ray spectrum is unusual for an SFXT, but resembles
  the so-called ``highly obscured sources'', high mass X-ray binaries (HMXBs) hosting
  an evolved B[e] supergiant companion (\sgb). 
  This might suggest that \src\ is a new member of this rare type of HMXBs, which
  includes IGR~J16318-4848 and CI Camelopardalis.
Increasing this small population of sources would be remarkable,
as they represent an interesting short transition evolutionary stage in the evolution of massive binaries.
Nevertheless,  \src\ appears to share X-ray properties of both
kinds of HMXBs (SFXT vs \sgb\ HMXB). Therefore, further investigations of the companion star
are needed to disentangle the two hypothesis.
\end{abstract}



\section{Introduction}

One of the legacies of every X-ray mission is the production of source catalogues. 
They always include a significant fraction of unidentified sources, whose nature remains unknown or controversial for
several years, awaiting further investigations.
This is the case of \src, a source discovered during ASCA observations (performed in 1996) of the  
Galactic supernova remnant (SNR) \rxj\ \citep{Uchiyama2002},
a shell-like SNR site of production of synchrotron X-ray emission \citep{Koyama1997, Slane1999}.
\src\ is located beyond the northern rim of the shell,
and at first it was suggested to be associated with a molecular cloud \citep{Uchiyama2002}.
The ASCA spectrum is well modeled by
a hard powerlaw  with a photon index $\Gamma$=0.98$^{+0.44} _{-0.34}$ and
an absorbing column density \nh=$1.28^{+1.00} _{-0.70}$ $\times$10$^{22}$~cm$^{-2}$.  
This absorption was very similar to the one measured from other regions of the SNR,  
consistent with the total Galactic one in the source direction
(\nh=1.5$\times$10$^{22}$~cm$^{-2}$; \citealt{nhcol2016}).  
The absorption-corrected flux was 4$\times10^{-11}$~\ferg (1-10 keV).
Given the spatial overlap of the ASCA source with the cloud, \citet{Uchiyama2002} interpreted the flat X-ray spectrum as 
produced by non-thermal bremsstrahlung from  particles accelerated in the SNR, then impacting the molecular gas as a target.
This hypothesis assumes the physical proximity of the SNR with the molecular cloud, and that \src\ is an extended X-ray source.

An observation performed in 2015 with the \chandra\ telescope proved the point-like character of \src\
(named \cxo, \citealt{Miceli2018}), 
excluding the association with the molecular cloud.
A relatively low X-ray flux was observed, 7($\pm{3})\times$10$^{-14}$~\ferg (2-10 keV; corrected for the absorption).
\citet{Miceli2018} reported also on a \suz\ observation of the field containing \src\ performed in 2011, where
the source showed a variable X-ray emission, with a short ($\sim$2~ks) hard X-ray flare.
The separate spectroscopy of the quiescent emission and the flare resulted into a quite high absorption 
(\nh\ in the range 3.6-12.2 $\times$10$^{22}$~cm$^{-2}$ for emission in quiescence, 
and \nh=6$-$11$\times$10$^{22}$~cm$^{-2}$ during the flare) and
flat power law spectra (photon index, $\Gamma$, in the range 0.6-2.3 
during quiescence, and $\Gamma$ ranging from 0.7 to 1.6 during the flare, at 90\% confidence level).  
The observed (not corrected for the absorption) fluxes (2-10 keV) were F=6$\times$10$^{-13}$~\ferg\ and
F=3.6$\times$10$^{-12}$~\ferg\ for the quiescent  and flare emission, respectively.

On this basis, \citet{Miceli2018} proposed that
\src\ is a high mass X-ray binary (HMXB) belonging to the sub-class of the Supergiant Fast X-ray Transients 
(SFXTs; \citealt{Sguera2005, Sguera2006, Negueruela2005a}).
The positional overlap with the 
near-infrared (NIR) point source 2MASS~17134391-3912055 
further supported the identification with
a massive X-ray binary \citep{Miceli2018}.

\begin{figure}
\includegraphics[width=\columnwidth]{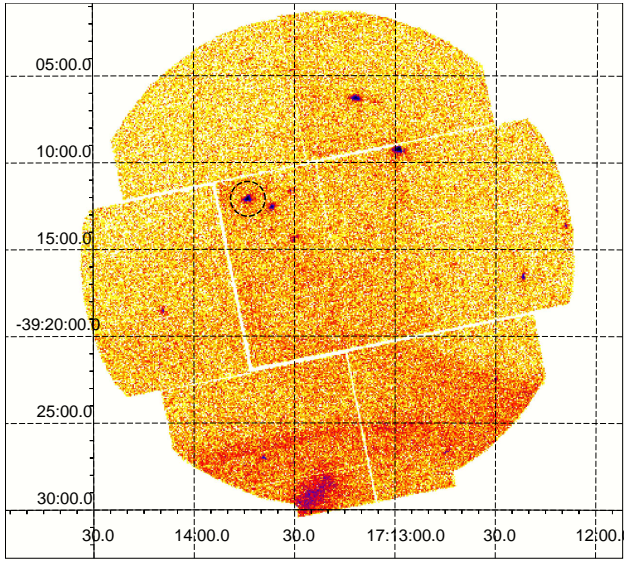}
    \caption{EPIC MOS2 image of the \xmm\ observation of the northern region of the SNR \rxj. 
      The sky position of \src\ is marked with the black, dashed circle.
      The northern rim of the SNR shell is evident
    in the lowest part of the image.}
    \label{fig:image}
\end{figure}

\section{Observations and Data Analysis}
\label{sect:data}

The sky position of  \src\  was serendipitously covered  by \xmm\ \citep{Jansen2001}
during an observation performed in 2017, from 29 August (at 15:28, UTC) to
30 August  (at 04:08), with an on-time exposure of 41.6\,ks (\pn) and 45.5\,ks (MOS).
The observation (Obs.ID 0804300901) was targeted at the northern region of the SNR \rxj\ and 
imaged \src\  at an offaxis angle of about 5 arcmin.
In Fig.\,\ref{fig:image} we show the \mosdue\ field-of-view (FOV), where \src\
is marked by a dashed black circle.

The three European Photon Imaging Cameras (\epic) \citep{Struder2001, Turner2001}
operated  with the medium filter, with the pn in full frame extended window, and
the two MOS in full frame mode.
EPIC data were reprocessed using the version 18 
of the \xmm\ Science Analysis Software (SAS), with standard procedures.
The tools {\em rmfgen} and {\em arfgen}, available in the SAS, were used to generate
the response and ancillary matrices, respectively.
High background levels were filtered-out before extracting EPIC spectra.
Light curves and spectra were extracted from circles centered on the source emission,
adopting a 30\arcsec\ radius, selecting patterns from 0 to 4 (\epic\ \pn), and from 0 to 12 (\mos).
Similar size regions, offset from the source position but lying on the same CCD, were used to
extract background spectra.
Source spectra from the \pn, \mosuno\ and \mosdue\  were simultaneously fitted
using {\sc xspec} (version 12.10.1; \citealt{Arnaud1996}) in the energy range 0.3-12 keV,
allowing for free cross-calibration constants,
to take into account calibration uncertainties.
All fluxes were estimated in the 1-10 keV range.
The models {\sc TBabs} and {\sc TBpcf} were adopted to account
for the absorbing column density along the line of sight,
assuming the photoelectric absorption cross sections of \cite{Verner1996}
and the interstellar abundances of \cite{Wilms2000}. 
The spectra were rebinned to have at least 20 counts per bin, to apply the $\chi^{2}$ statistics. 
All uncertanties are computed at 90\% confidence level, for one interesting parameter.
The uncertainty on the  X-ray fluxes have been calculated using {\sc cflux} in {\sc xspec}.

\section{Results}
\label{sect:res}

\subsection{Temporal analysis}
\label{sect:timing}

\begin{figure}
    \includegraphics[width=10.0cm, height=8.5cm, angle=-90]{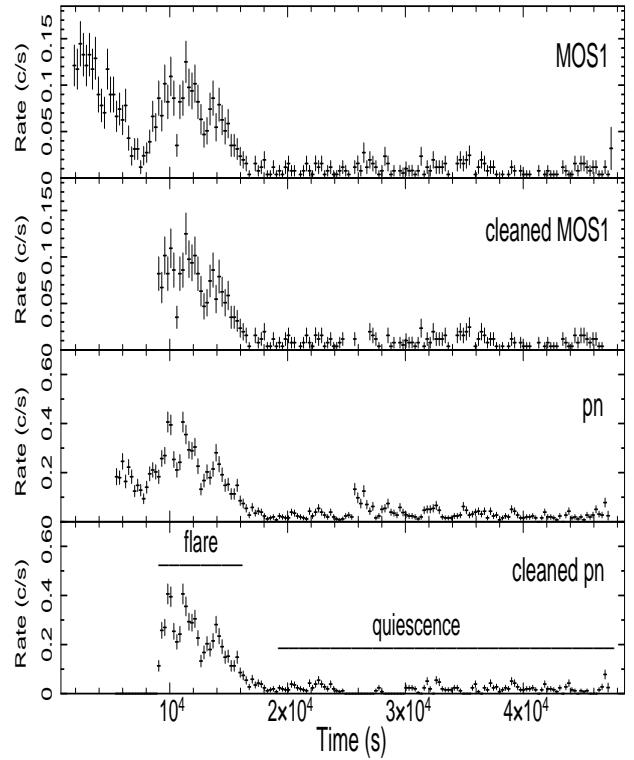}  
\caption{\src\ light curve in the energy band 0.3-12 keV (bin time = 256 s)  observed by \xmm.
  From top to bottom,  we show the MOS1 source light curve (MOS2 one 
  is similar), the cleaned MOS1 one (where time intervals with
  high background levels have been filtered-out), the EPIC pn light curve
  and the cleaned one (lowest panel). 
  The horizontal lines in the lowest panel indicate the time intervals for
  the extraction of the flare and quiescent EPIC spectra (the same for pn,  MOS1 and MOS2).
}
    \label{fig:lc}
\end{figure}

The EPIC source  light curve is reported in Fig.\,\ref{fig:lc},  before and after filtering for high background levels.
The source 
displays a significant variability on timescales of a few hundred seconds,
with a dynamic range of $\sim$50.
Although formally the energy range is 0.3-12 keV, we note that most
of the source counts lies in the 2-12 keV energy band.
We have identified two source states: a high state (or ``flare'', hereafter) and a low one, named ``quiescence'', indicated
in the lowest panel in Fig.\,\ref{fig:lc}.

We searched the data for periodic signals by means of  Fourier transforms and Rayleigh periodograms, but we did not find any statistically significant signal. The 3$\sigma$ upper limit on the pulsed fraction, computed by extensive Monte Carlo simulations, is 25\% for a sinusoidal signal between 0.4 and 5.4\,s, using only the pn data (2-12 keV), and 20\% between 5.4 and 1000\,s, using also the data from the MOS cameras. Above $\approx$1000 s, the strong red noise does not allow us to set meaningful limits.

\subsection{Spectroscopy}
\label{sect:spec}

\begin{figure}
\includegraphics[width=12.0cm, height=8.5cm, angle=-90]{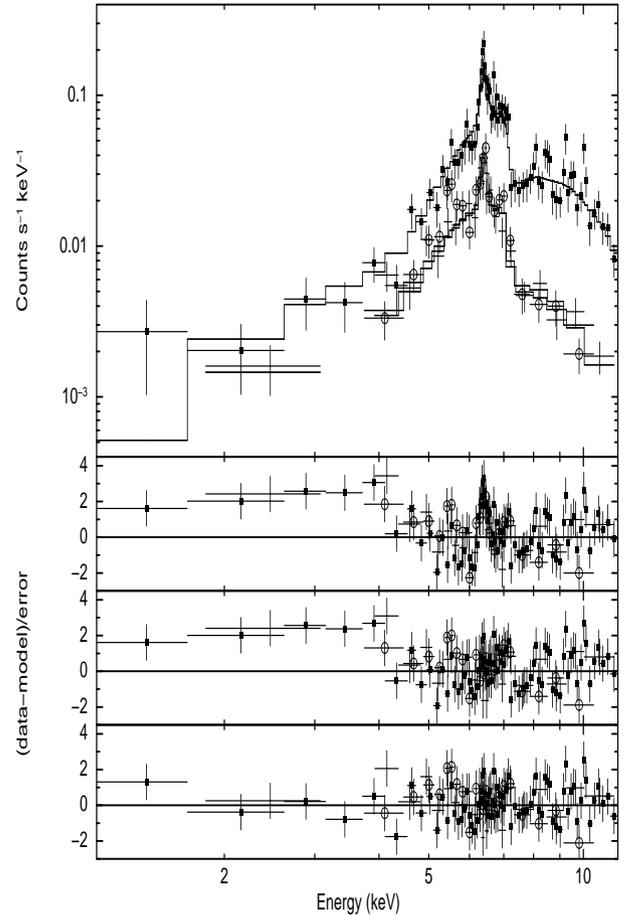}
\caption{EPIC spectra extracted during the flare.
  Counts spectra are shown in the top panel (EPIC \pn\ is marked with solid squares, \mosuno\ with open circles, \mosdue\ with crosses), when fitted with the best fit reported in Table\ref{tab:spec}.
  Residuals (in units of standard deviation) with respect to three models
  are shown in the bottom panel: from top to bottom, the residuals are with respect to a single absorbed powerlaw, with respect to a powerlaw with a Gaussian line at 6.4 keV, and including a partial covering absorption model (i.e. the best fit reported in Table\ref{tab:spec}).
}
    \label{fig:spec_flare}
\end{figure}

We performed the spectroscopic analysis during the flare and in quiescence separately, extracting two sets of EPIC spectra,
  covering the time intervals shown in Fig.\,\ref{fig:lc} (bottom panel).
The EPIC spectra extracted during the flare are highly absorbed and with large
positive residuals around 6.4 keV,
very evident when fitting the spectra with a simple, absorbed power law model
(Fig.\,\ref{fig:spec_flare}), resulting in a reduced $\chi^{2}_{\nu}$=1.928 (for 117 degrees of freedom, dof;
 see Table\,\ref{tab:spec} for the spectral parameters).
The addition of a narrow Gaussian line accounted for these residuals
($\chi^{2}_{\nu}$/dof=1.445/114;  Table\,\ref{tab:spec}).
However, a mild soft excess remained below 4 keV.
This suggested to consider an additional absorption model, a partial covering fraction absorption
({\sc TBpcf} in {\sc xspec}), where the additional column density is applied  to a fraction of the power law emission.
This final model provides a  good description of the flare spectrum (Model 3 in Table\,\ref{tab:spec}).
The counts spectra and the residuals to this best fit model are shown in Fig.\,\ref{fig:spec_flare}.

We note that the model {\sc TBpcf} resulted into an almost complete covering (98$\pm{1}$\%) of the X-ray emission,
leading to an absorption of 1.5$\times$10$^{24}$\,cm$^{-2}$ during the flare, considering both absorbing components
({\sc TBabs} and {\sc TBpcf}). This is 2 dex larger than the Galactic absorption (\nh=1.5$\times$10$^{22}$~cm$^{-2}$; \citealt{nhcol2016}). 
The centroid of the emission line is very well constrained in a narrow range around
6.4 keV in both states, clearly indicative of fluorescence from neutral iron.

The Fe K$_{\alpha}$ line is produced when a direct hard X-ray radiation
illuminates neutral matter around the source.
This reprocessing results into a fluorescent iron line emission together with
a Compton component.
The final spectrum is a composition of a
reflection component together with the direct (power law) X-ray emission.
However, adopting reflection models like {\sc pexrav} \citep{pexrav} and 
{\sc pexmon} \citep{Nandra2007} in {\sc xspec} did not yield  better fits.
We note that disentangling the incident and the reflected component is 
made difficult due to the limited energy band and the low counting statistics.
This results into a quite hard power law slope measured in the total spectrum below 10 keV,
  also because of a reflected hump  \citep{Fabian1990} that is expected beyond the \xmm\ energy band.

\begin{figure}
   \includegraphics[width=9.0cm, height=8.5cm, angle=-90]{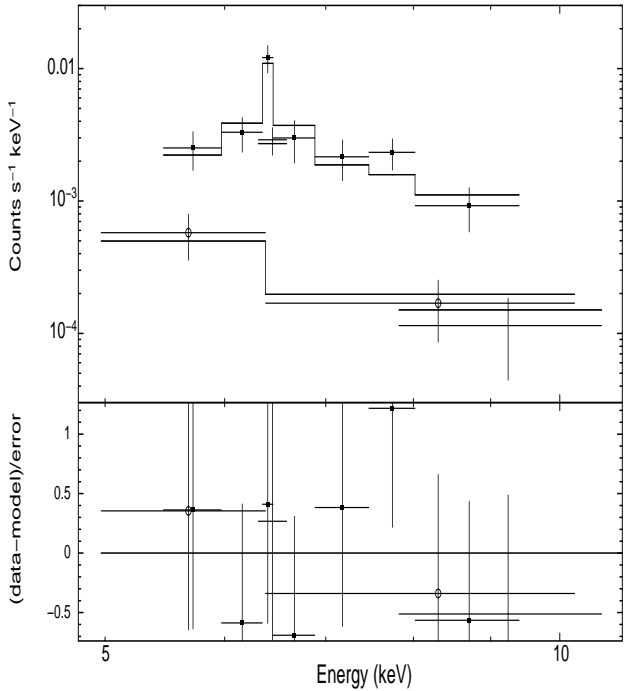}
\caption{EPIC spectra extracted during  quiescence.
  Counts spectra are shown in the top panel (EPIC \pn\ is marked with solid squares, \mosuno\ with open circles, \mosdue\ with crosses), when fitted with the best fit reported in Table\ref{tab:spec}, an absorbed power law together with an emission line at 6.4 keV. The  lower panel shows the residuals  in units of standard deviation.
}
    \label{fig:spec_quiesc}
\end{figure}

The spectrum in quiescence shows a prominent positive excess around 6.4 keV as well,
when fitted with a simple absorbed power law
($\chi^{2}_{\nu}$/dof=2.85/6). 
The addition of a Gaussian line to the absorbed
power law model resulted in a good fit to the data, with no need to adopt
further partial covering  absorption (Fig.\,\ref{fig:spec_quiesc}).
During the fit, a narrow iron emission line is assumed, fixing its width to zero.
The best fit parameters are listed in  Table\,\ref{tab:spec} (last column).

\begin{table*}
  \caption{Spectroscopy of the two source states (\epic\ \pn, \mosuno\ and \mosdue).
    We show spectral results obtained using three models for the flare emission: Model 1 is a simple absorbed power law, Model 2 includes an iron line in emission, while
    the best fit model is  Model 3, a partially absorbed powerlaw, together with a Gaussian line in emission ({\sc  const * TBabs * TBpcf * (POW+GAU)} in {\sc xspec}  synthax).
     The last column lists the best fit parameters for the spectroscopy during quiescence ({\sc  const * TBabs * (POW+GAU)}). 
F$_{1-10 keV}$ is the absorbed flux, UF$_{1-10 keV}$ the flux corrected for the absorption.
  }
\label{tab:spec}
\vspace{0.0 cm}
\begin{center}
\begin{tabular}{lcccc} \hline
 \hline
\noalign {\smallskip}
Parameters                               &   \multicolumn{3}{c}{Flare}                                           &   Quiescence     \\
                                      &   Model 1                   & Model 2             & Model 3                 &                  \\
\hline
\noalign {\smallskip}
N$_{\rm H}$ (10$^{22}$ cm$^{-2}$)          & $127^{+16} _{-15}$         & $99^{+18} _{-17}$    & $5.8^{+6.3} _{-4.0}$      &      $< 80$    \\
\multicolumn{2}{c}{----- Partial covering fraction absorption -------}   \\
N$_{\rm H TBpcf}$ (10$^{22}$ cm$^{-2}$)     &     $-$                  &     $-$         &  $145\pm{20}$                 &     $-$   \\
covering fraction                        &     $-$                  &     $-$         &  $98\pm{1}\%$                 &     $-$  \\
\multicolumn{2}{c}{----- POWER LAW-------}   \\
$\Gamma$                         & $0.21^{+0.32} _{-0.32}$             & $-0.40^{+0.39} _{-0.39}$             &    $0.06^{+0.37} _{-0.38}$             &  $-0.8^{+1.6} _{-1.7}$ \\
norm      	                 & $10(^{+12} _{-6})\times10^{-4}$     & $2.0(^{+3.3} _{-1.2})\times10^{-4}$  &   $8.4(^{+13.5}_{-5.3})\times10^{-4}$   &  $1.1(^{+50.}_{-1.0})\times10^{-6}$  \\
\multicolumn{2}{c}{----- GAUSSIAN LINE-------}   \\
E$_{\rm line}$   (keV)                   &     $-$     &    $6.409^{+0.025} _{-0.027}$         &  $6.402^{+0.024} _{-0.026}$           &    $6.41^{+0.06} _{-0.05}$              \\
$\sigma$  (keV)                         &     $-$    &   $<$0.10	                   &  $<$0.08	                          &       0.0 (fixed)	                    \\
norm  (photons cm$^{-2}$ s$^{-1}$)       &     $-$     &   $1.3^{+0.4} _{-0.3} \times10^{-4}$  &  $2.0^{+0.7} _{-0.5} \times10^{-4}$   &   $4.4^{+6.3} _{-1.8}  \times10^{-6}$       \\
EW   (eV)                               &     $-$     &   $320^{+80} _{-70}$                 &  $265^{+30} _{-80}$                   &        $900^{+1300} _{-400}$    	\\
\hline
F$_{1-10 keV}$   (\flux)        & 9.2 $\pm{0.5}$ $\times10^{-12}$        & 9.3 $\pm{0.5}$ $\times10^{-12}$           &   9.4 $\pm{0.5}$ $\times10^{-12}$      & 4.3$^{+0.9} _{-1.2}$ $\times10^{-13}$   \\
UF$_{1-10 keV}$   (\flux)       & 5.4 $^{+1.6} _{-1.2}$ $\times10^{-11}$  & 3.4 $^{+1.1} _{-0.8}$ $\times10^{-11}$      &   6.2 $^{+2.5} _{-1.7}$ $\times10^{-11}$ &  4.4$^{+3.1} _{-0.1}$ $\times10^{-13}$    \\
\hline
L$_{1-10 keV}$  (\ergsec)       &    6.5$\times10^{35}$ d$_{10kpc}^2$   &    4.1$\times10^{35}$ d$_{10kpc}^2$         &    7.4$\times10^{35}$ d$_{10kpc}^2$      &  5.3$\times10^{33}$ d$_{10kpc}^2$    \\
\hline
$\chi^{2}_{\nu}$/dof              &   1.928/117    &   1.445/114  &   1.076/112                        &    0.854/4 	 \\
\hline
\hline
\end{tabular}
\end{center}
\end{table*}

\section{Discussion}

We have reported  on the discovery of a  prominent FeK$\alpha$  line
and a large intrinsic absorption (\nh=1.5$\times$10$^{24}$\,cm$^{-2}$)
from \src\ during an \xmm\ observation performed in 2017.
The source  shows a variable X-ray emission with a brighter state (flare) at the beginning of the observation,
followed by a fainter state (quiescence).
Spectra extracted from both states are well described by hard power law models with similar slopes,
within the uncertainties.
The X-ray emission is significantly more absorbed during the flare than during the following fainter state,
suggesting a variability in the circumstellar absorbing matter on short timescale, correlated with the
X-ray flux.

The FeK$\alpha$ is detected in both states, with a  larger equivalent width during the quiescent emission.
This might suggest that the unflared state is due to the eclipse by the companion star: the
less absorbed X-ray spectrum can be due to scattering  into the line of sight by the stellar wind matter
of the central (eclipsed) X-ray radiation (e.g. \citealt{Haberl1991}).
However, the flux of the iron line is significantly different in the two states and
correlates with the continuum flux, disfavouring this hypothesis, and suggests an intrinsic variability.
Moreover, this correlation indicates  a close
proximity of the reprocessing matter to the compact object.

\begin{figure}
  \includegraphics[height=\columnwidth, angle=-90]{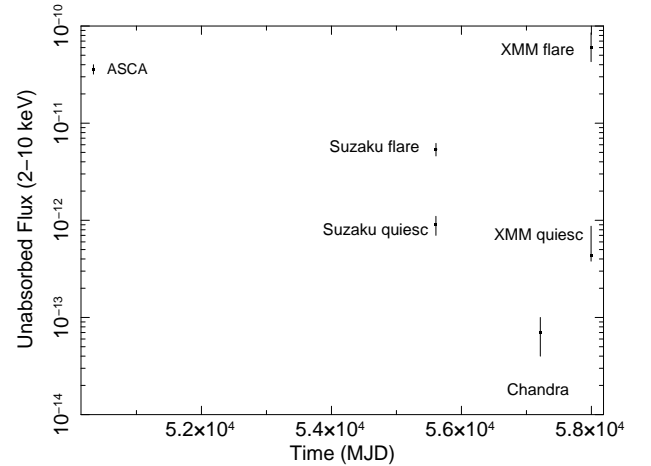}
  \caption{\src\ long term light curve (fluxes in the energy range 2-10 keV are corrected for the absorption).
}
    \label{fig:flux}
\end{figure}

The source is also variable on long timescales of years:
previous X-ray observations (ASCA in 1996, \suz\ in 2011, \chandra\ in 2015)
caught different X-ray fluxes (Fig.\,\ref{fig:flux})
and a significantly lower absorbing column density
(\nh\ in the range 10$^{22}$-10$^{23}$\,cm$^{-2}$) than during the \xmm\ observation.
This indicates a long-term changing aspect of the absorbing matter local to the source,
possibly with an inhomogeneous distribution or a variability due to the orbital motion
in a binary system. 

In light of the new \xmm\ results, we discuss the possible source nature
in the following sub-sections.

\begin{figure}
  \includegraphics[width=\columnwidth, angle=0]{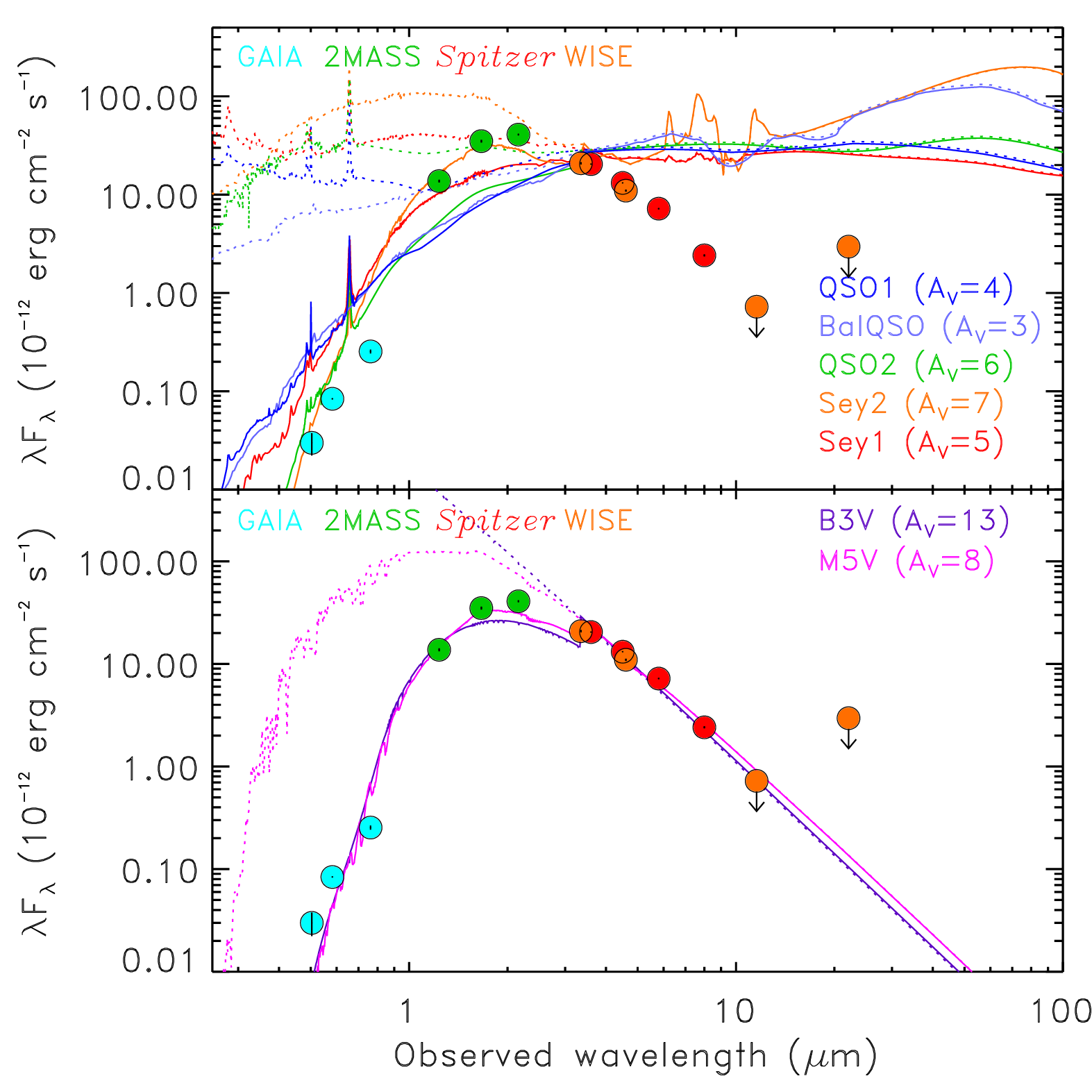}
\caption{Spectral energy distribution of AX\,J1714.1$-$3912 (full circles: GAIA in
cyan, 2MASS in green, \textit{Spitzer} in red and WISE in orange).  The
lines represent various templates normalized at the observed 3.6$\mu$m flux
with no reddening (dotted) or with different amounts of reddening (solid),
as annotated, applied to reproduce the observed optical-NIR SED.  Templates
of various types of AGN are shown in the \textit{top
panel}~\citep{polletta07}, and of a B and M-type star~\citep{kurucz93} in
the \textit{bottom panel}.
}
    \label{fig:sed}
\end{figure}

\subsection{Is \src\ an active galactic nucleus?}

The point-like appearance, the large intrinsic absorption and the presence
of an FeK$\alpha$ line in AXJ\,1714.1$-$3912 are reminiscent of the X-ray
properties of obscured active galactic nuclei~\citep[AGN; see
e.g.,][]{guainazzi05}. To investigate such a hypothesis we examine the
broad-band spectral energy distribution (SED) of the optical-infrared (IR)
counterpart to AXJ\,1714.1$-$3912.  To build the broadband SED we collected
optical data from GAIA DR3~\citep{gaia16_mission,gaia21_dr3}, near-IR data
from 2MASS~\citep{skrutskie06}, and mid-IR data from WISE~\citep{wright10}
and $Spitzer$~\citep{werner04,glimpse}.  The optical-IR SED, shown in
Fig.~\ref{fig:sed}, peaks at $\sim$2$\mu$m and decreases steadily towards
longer wavelengths up to $\lambda{\leq}$10$\mu$m.  Such a behaviour is
inconsistent with what is observed in AGN, whose SEDs typically rise
long-ward of 1--5$\mu$m~\citep{polletta07,hickox17} due to the emission from AGN-heated
hot dust.
Therefore, the broad-band SED rules out the AGN hypothesis.
The mid-IR ($\lambda$\,=\,3--10$\mu$m) SED of AXJ\,1714.1$-$3912
is instead consistent with stellar radiation.  The full optical-IR SED of
AXJ\,1714.1$-$3912 can be reproduced with various reddened stellar
templates~\citep{kurucz93}. We apply a standard Galactic reddening
law~\citep{cardelli89}. The amount of required optical extinction
depends on the stellar type, for example an A$_\mathrm{V}$ of 13\,mag,
corresponding to N$_\mathrm{H}{\simeq}$2.3$\times$10$^{22}$\,cm$^{-2}$, is
required for a B-type star, and an A$_\mathrm{V}$ of 8\,mag
(N$_\mathrm{H}{\simeq}$1.4$\times$10$^{22}$\,cm$^{-2}$) for a red-giant
M-type star.  The estimated column densities are consistent or slightly
larger than the Galactic value measured along the line of sight towards
AXJ\,1714.1$-$3912~\citep[i.e.,
  1.5$\times$10$^{22}$\,cm$^{-2}$;][]{nhcol2016}.  
Spectroscopic observations
and a more detailed analysis would be necessary to better characterise the
stellar type and determine whether intrinsic dust absorption is present in
AXJ\,1714.1$-$3912.

\subsection{Is \src\ an SFXT?}

\citet{Miceli2018} excluded an extragalactic origin as well,
based on the rapid timescale (thousands seconds) of the X-ray flux variability.
They suggested that \src\ is a Galactic HMXB belonging to the sub-class of SFXTs,
based on the point-like appearence, the amplitude of the X-ray variability
and the hard power law spectrum, indicative of accretion of matter onto a compact object.

In light of the flare caught by \xmm, the source dynamic range (F$_{\rm max}$/F$_{\rm min}$) increases to 
$\sim$900, compared with the faint flux detected by \chandra\ (Fig.\,\ref{fig:flux}).
This range of variability is not as extreme as the one shown by the prototypical members of the SFXT class
(F$_{\rm max}$/F$_{\rm min}$ from 10$^{4}$ to 10$^{6}$) but still consistent with 
less variable, ``intermediate'' SFXTs (see Table\,2 in \citealt{Sidoli2018}).
However it is possible that we missed the brightest flares, since the duty cycle of SFXT outbursts is
very small (lower than 5\%, see Table\,1 in \citealt{Sidoli2018}).
On the other hand, the source field has been monitored by \inte/IBIS (above 20 keV) for a total
exposure time of 6.7\,Ms \citep{Bird2016}, with no detections reported in the literature.
If we assume a typical flare duration of $\sim$2\,ks  we calculate a
duty cycle lower than 0.03\% for \src\  (percentage of time spent in bright flaring activity, i.e.
with a peak flux F$_{18-50 keV}\geq$1.5-3$\times$10$^{-10}$~\flux; \citealt{Sidoli2018}),
to reconcile with the lack of reported outbursts with \inte.
This would imply that \src\ is the SFXT with the lowest duty cycle:
to date, the SFXT with the rarest outbursts is IGR\,J08408-4503,
showing a duty cycle of 0.09\% \citep{Sidoli2018}.
Alternatively, \src\ could be located at large distance: a short (duration $\sim$2\,ks) flare
with a peak luminosity of $\sim$10$^{36}$~\ergsec\ (18-50 keV)
implies a source distance d$\gtrsim$8~kpc, to not be detected by \inte.

The non detection with \inte\ poses also a
3$\sigma$ upper limit to the persistent (quiescent) emission
F$<$2.3$\times10^{-12}$\,\flux\ (20-40 keV; \citealt{Bird2016}).
If we assume  that the quiescent \xmm\ spectrum
  is representative of the long-term source state, we can use this upper limit to constrain the presence of a high energy cutoff,
  despite  the large uncertainty in the measured power law slope. 
  In particular, if the true photon index of the quiescent spectrum is $\Gamma$$\sim$0.8, the extrapolation of the power law model
at higher energies leads to F$=$1.3$\times10^{-12}$\,\flux\ (20-40 keV), with no need for a cutoff.
If  harder power law photon indexes are assumed, variable cutoff values are
needed to reconcile with the \inte\ upper limit.
For instance, for a photon index $\Gamma$$\sim$$-1$ in the quiescent spectrum, a cutoff E$_{cut}$$\sim$10 keV is needed,
to match the upper limit to the 20-40 keV flux (where E$_{cut}$ is the  e-folding energy of exponential 
rolloff in the {\sc cutoffpl} model in {\sc xspec}).

\subsection{Is \src\ a supergiant B[e] HMXB?}

Although the long-term X-ray light curve is compatible with an SFXT with very 
rare outbursts (and/or located at large distance),
we note that the \xmm\ spectral properties reported here for the first time
are unusual for an SFXT \citep{Sidoli2017review, Martinez2017, Kretschmar2019}:
so far,  no members of this class are known to be so highly absorbed.
Even in the SFXT IGR\,J18410$-$0535,  where an intense flare was suggested to be produced by accretion
of a very massive wind clump, the associated
absorbing column density was significantly lower \citep{Bozzo2011}.
The most extreme absorption among SFXTs has been observed in SAX\,J1818.6$-$1703
(5$\times$10$^{23}$\,cm$^{-2}$; \citealt{Boon2016}), but it is a unique case.
In general, SFXTs show a circumstellar environment less dense than in persistent HMXBs \citep{Kretschmar2019}.
We note that large absorbing column densities, variable on timescales of ten days, have been measured in the 
Be X-ray transient SXP\,1062 \citep{Gonzalez2018} during the decline of the outburst,
but with no detection of FeK$\alpha$ line emission.

On the other hand, the \src\ spectrum observed with \xmm\ strongly resembles those of
the so-called ``highly obscured sources'' \citep{Walter2003}.
The latter are HMXBs where the compact object is enshrouded in a dense circumstellar
environment produced by the outflowing matter from an evolved, early type massive star, such as a \sgb.
In particular, the huge absorbing column density of $\sim$1.5$\times$10$^{24}$\,cm$^{-2}$ we deduce from the
flare emission, makes \src\ one of the most absorbed sources ever observed in our Galaxy, together
with IGR J16318-4848 \citep{Ibarra2007}. 

SgB[e] stars \citep{Zickgraf1985, Kraus2019} are evolved massive stars characterized by disk-like,
dusty circumstellar envelopes fed by dense outflows from the B supergiant.
Their optical spectra show a twofold behaviour: broad Balmer emission lines 
plus narrow emission lines from permitted and forbidden transitions. 
HMXBs with a supergiant B[e] donor stars are a rare type of X-ray binaries,
with CI Camelopardalis (CI Cam, aka XTE\,J0421+560; \citealt{Bartlett2013})
as a prototype, being the first Galactic sgB[e] star observed during an X-ray outburst
whose X-ray luminosity implied a clear binarity nature.
\citet{Chaty2019} suggest that \sgb\ HMXBs are at the short evolutionary stage when
a binary system is entering a common envelope phase of binary evolution.
At present, this is a small class of rare HMXBs that, besides 
the Galactic sources
CI\,Cam, IGR J16318$-$4848, and Wd1-9, includes a couple of
candidates in the Magellanic Clouds and, remarkably, two ultra luminous X-ray sources,
Holmberg II X-1 and NGC300 ULX-1/supernova imposter SN2010da \citep{Bartlett2019}.

CI Cam and IGR J16318$-$4848 show variable absorbing column densities in the range 10$^{23}$-10$^{24}$\,cm$^{-2}$,
intense FeK$\alpha$ line emission and X-ray flux variability \citep{Bartlett2019}.
But while IGR J16318-4848 is bright above 20 keV \citep{Bird2016} with some level of flaring activity \citep{Sidoli2018},
CI Cam has never been detected by \inte\ \citep{Bird2016}.
Nevertheless, IGR J16318$-$4848 has never undergone an X-ray outburst similar to the one
experienced by CI Cam in 1998: CI Cam displayed a dynamic range
in excess  of 500 in 8 days (from $\sim$5$\times$10$^{-8}$\,\flux\ to 9$\times$10$^{-11}$\,\flux),
with a decline of five
orders of magnitude, back to quiescence, in a few months \citep{Belloni1999, Orlandini2000}.
Their X-ray luminosities cannot be determined as their distances are very uncertain.
The nature of the compact object is unknown as X-ray pulsations have not been observed.

\section{Conclusions}

 We have reported here on an \xmm\ observation of \src, leading to the discovery of a remarkable
  new behavior of this source. The new findings can be summarized as follows:

\begin{itemize}
\item   a high intrinsic obscuration ($\sim$1.5$\times$10$^{24}$\,cm$^{-2}$)
  is observed during the flaring emission, implying a large variability of two orders of magnitude in
  the absorbing column density towards the source, on timescales of years;
  
\item a prominent FeK$\alpha$ line emission is evident both during
  the flare and the unflared (quiescent) emission, with variable fluxes in the two source states;

\item the flare caught by \xmm\ increases the source range of flux variability to $\sim$900,
  when compared to a \chandra\ observation performed two years before.
  
\end{itemize}

In view of these new findings we have discussed different viable scenarios for the source nature.
\src\ was previously suggested to be a SFXT.
The short term variability during the \xmm\ observation is consistent with a SFXT nature,
as well as the long-term dynamic range.

On the other hand, the  \xmm\  spectrum is remarkable,
as no SFXT has ever shown an obscuration as large as  10$^{24}$\,cm$^{-2}$.
This spectrum shows many similarities with those typically observed in
the so-called ``highly obscured sources'',
a rare sub-class of HMXBs with a \sgb\ companion.
This might pose the SFXT identification into question and leads us to propose an alternative origin
for the X--ray emisison, a \sgb\ HMXB.
To confirm its membership further investigations of the companion star are needed.

\section*{Acknowledgements}

Based on observations (ObsID 0804300901) obtained with \xmm, a European Space Agency science mission with instruments and contributions
directly funded by ESA Member States and NASA.
This work has made use of data and software provided by the High Energy Astrophysics Science Archive Research Center (HEASARC), 
which is a service of the Astrophysics Science Division at NASA/GSFC.
This work has made use of data from the ESA mission
{\it Gaia} (\url{https://www.cosmos.esa.int/gaia}), processed by the {\it Gaia}
Data Processing and Analysis Consortium (DPAC,
\url{https://www.cosmos.esa.int/web/gaia/dpac/consortium}). Funding for the DPAC
has been provided by national institutions, in particular the institutions
participating in the {\it Gaia} Multilateral Agreement.
This publication makes use of data products from the Two Micron All Sky
Survey, the Wide-field Infrared Survey Explorer, and \textit{Spitzer} Space Telescope.
2MASS is a joint project of the University of Massachusetts and the
Infrared Processing and Analysis Center/California Institute of Technology,
funded by the National Aeronautics and Space Administration and the National
Science Foundation.
WISE  is a joint project of the University of California,
Los Angeles, and the Jet Propulsion Laboratory/California Institute of
Technology, funded by the National Aeronautics and Space Administration.
\textit{Spitzer}  was operated by the Jet Propulsion
Laboratory, California Institute of Technology under a contract with NASA.
PE acknowledges financial support from the Italian Ministry for Education and Research through the PRIN grant 2017LJ39LM.

\section*{Data Availability}

The \xmm\ data analysed here are publicly available by means of the HEASARC  (ObsID 0804300901) 
and the ESA archive at the link https://doi.org/10.5270/esa-nai97jb


\bibliographystyle{mnras}

\bsp	
\label{lastpage}
\end{document}